\documentclass[prd,tightenlines,nofootinbib,superscriptaddress]{revtex4}

\usepackage{amsfonts,amsmath,amssymb,amsthm,bbm,hyperref}
\usepackage{color,psfrag}
\usepackage[dvips]{graphicx}

\newcommand{\R}{{\mathbb R}}

\newcommand{\be}{\begin{equation}}
\newcommand{\ee}{\end{equation}}
\newcommand{\beq}{\begin{eqnarray}}
\newcommand{\eeq}{\end{eqnarray}}
\newcommand{\bea}{\begin{eqnarray}}
\newcommand{\eea}{\end{eqnarray}}
\newcommand{\nn}{\nonumber}

\newcommand{\su}{{\mathfrak su}}

\newcommand{\ra}{\rangle}

\def\nn{\nonumber}

\begin{document}

\title{Revisiting the quantum scalar field in spherically symmetric
quantum gravity}

\author{{\bf Enrique F. Borja}}
\affiliation{Institute for Theoretical Physics III, University of
Erlangen-N\"{u}rnberg, Staudtstra{\ss}e 7, D-91058 Erlangen (Germany)}
\affiliation{Departamento de F\'{\i}sica Te\'{o}rica and IFIC, Centro Mixto
Universidad de Valencia-CSIC. Facultad de F\'{\i}sica, Universidad de
Valencia, Burjassot-46100, Valencia (Spain)}
\author{{\bf I\~naki Garay}}
\affiliation{Institute for Theoretical Physics III, University of
Erlangen-N\"{u}rnberg, Staudtstra{\ss}e 7, D-91058 Erlangen (Germany)}
\affiliation{Departamento de F\'{\i}sica Te\'{o}rica, Universidad del Pa\'{\i}s
Vasco, Apartado 644, 48080 Bilbao (Spain)}
\author{{\bf Eckhard Strobel}\email{eckhard@bebsdb.dnsalias.net}}
\affiliation{Institute for Theoretical Physics III, University of
Erlangen-N\"{u}rnberg, Staudtstra{\ss}e 7, D-91058 Erlangen (Germany)}

\date{\today}

\begin{abstract}
We extend previous results in spherically symmetric gravitational
systems coupled with a massless scalar field within the loop quantum
gravity framework. As starting point, we take the Schwarzschild
spacetime. The results presented here rely on the uniform
discretization method. We are able to minimize the associated
discrete master constraint using a variational method. The trial
state for the vacuum consists of a direct product of a Fock vacuum
for the matter part and a Gaussian centered around the classical
Schwarzschild solution. This paper follows the line of research
presented by Gambini, Pullin and Rastgoo \cite{Gambini09} and a
comparison between their result and the one given in this work is
made.
\end{abstract}

\maketitle


\tableofcontents

\section*{Introduction}

Loop quantum gravity (LQG) is a promising proposal for the canonical
quantization of general relativity \cite{Thiemann}. It provides a
non-perturbative and mathematically rigorous description of the
kinematical sector. Nevertheless, as it is well known, the treatment
of the dynamics of the theory remains an open problem. The main
difficulty of this task is due to the fact that the constraints in
general relativity do not satisfy a Lie algebra.

During the last years, there have been advances in the understanding
of the spherically symmetric gravity within the LQG framework, both
in the vacuum and in the case coupled with matter
\cite{Campiglia07,Gambini09}. Gambini, Pullin and Rastgoo studied
the problem of a gravitational system with spherical symmetry
coupled to a spherically symmetric massless scalar field in a
discretized space. They employed the uniform discretization method
\cite{Campiglia06,Campiglia06-2,Bahr:2011xs} to deal with the fact
that the Hamiltonian constraint satisfies a non-Lie Poisson bracket
with itself (after a gauge fixing of the diffeomorphism constraint).
This uniform discretizations technique provides a method to quantize
totally constrained systems. It is as the `master constraint'
program \cite{Thiemann:2003zv} an approach to find an alternative
for the Dirac quantization used in LQG.

More specifically, in \cite{Gambini09} the scalar field starting
from the Minkowski spacetime is studied. The authors performed a
minimization of the master constraint through a suitable variational
method. In this way, one can check whether the theory provides us
with the vacuum state of the scalar field centered around a
Minkowskian background. In the present paper we extend that work to
the Schwarzschild case. Although it is a modest objective, our
motivation is twofold. On the one hand, we want to test the
methodology presented by Gambini et al. in a less trivial case and,
on the other hand, we want to compare the final result for the
Schwarzschild spacetime with the one in Minkowski. As we are going
to show in the paper, although in this case the procedure becomes
more involved than the previous one, the final result is, at the
leading order, independent of the Schwarzschild radius.

This paper is organized as follows. In section \ref{sec:Uni} we
review the uniform discretizations method, that will be the arena
for our study. In section \ref{sec:Spher} we summarize the classical
treatment for spherical symmetric gravity coupled with a massless
scalar field. More concretely, we will write explicitly the
Hamiltonian constraint for this system. In the next section
\ref{sec:gravtrial}, we construct the trial state (composed by a
gravitational part quantized polymerically and the Fock vacuum for
the matter part) that we will use  later in section
\ref{sec:expectation} to evaluate and minimize the discrete master
constraint. In order to make the main text and ideas more clear, we
have gathered most of the technical calculations in the appendix and
we added the corresponding references where necessary. Of special
importance is the appendix \ref{app:SL}, where we comment the
solutions of the Sturm-Liouville equation that plays a central role
for the Fock quantization of the matter part.

\section{Uniform discretizations}
\label{sec:Uni}

The uniform discretizations technique
\cite{Campiglia06,Campiglia06-2,Bahr:2011xs} is a method to quantize
totally constrained systems. It is, as the master constraint program
\cite{Thiemann:2003zv}, an approach to find an alternative for the
Dirac quantization used in canonical LQG. Although LQG showed
remarkable results, especially concerning microscopic pictures of
the early universe and black holes, the dynamics of the full theory
still faces problems. One of those is that the quantum constraint
algebra fails to portray the classical algebra exactly. Among the
other options under research is the uniform discretization method.

The quantization of discretized, totally constrained systems faces
the problem that, in general, the constraint algebra fails to close,
i.e., the systems turn out to be second class even if the continuum
theory is first class. If one now uses the Dirac method to quantize
the discrete theory, the degrees of freedom will not match the ones
of the continuum. Furthermore, in general the symmetries of the
theory are not recovered and taking the continuum limit one will not
find the quantized version of the theory one started with.

If one has first class systems the uniform discretization technique
is akin to the Dirac procedure, so we concentrate on second class
systems here.

Starting with the study of a classical, totally constrained system
with $N$ configuration variables and $M$ constraints \(\phi_j\); we
consider the discretization of the constraints with a parameter
\(\epsilon\), such that \(\phi_j=\lim_{\epsilon\rightarrow0}
\phi_j^\epsilon\). As mentioned beforehand we assume the constraints
to be second class in the discrete theory:
\begin{equation}
 \left\{\phi_j^\epsilon,\phi_k^\epsilon\right\}
 =C_{jk}^{\epsilon \enspace m}\phi_m^\epsilon+A^\epsilon_{jk}\,.
 \label{eq:dca}
\end{equation}
In order to get first class constraints in the continuum theory we
impose \(\lim_{\epsilon\rightarrow 0}A^\epsilon_{jk}=0\)\,, and for
the structure functions of the continuum theory \(C_{jk}^{\enspace
\enspace m}=\lim_{\epsilon\rightarrow 0}C_{jk}^{\epsilon \enspace
m}\).

We now construct a discrete ``master constraint''
\begin{equation}
 \mathbb{H}^\epsilon=\frac{1}{2}\sum_j^M
 \left(\phi_j^\epsilon\right)^2\,.
 \label{eq:unimas}
\end{equation}
We proceed by introducing the time evolution for a discrete variable
\(A_n\) as
\begin{equation}
 A_{n+1}=e^{\left\{\,\cdot\,,\mathbb{H}^\epsilon\right\}}A_n
 :=A_n+\left\{A_n,\mathbb{H}^\epsilon\right\}+
 \frac12\left\{\left\{A_n,\mathbb{H}^\epsilon\right\},
 \mathbb{H}^\epsilon\right\}+\ldots
\end{equation}
It is obvious that \(\mathbb{H}^\epsilon\) is a constant of motion
under this time evolution. We fix its value to
$\mathbb{H}^\epsilon=\delta^2/2$ and define the quantities
$\lambda_i^\epsilon:=\phi_i^\epsilon/\delta$. With these definitions
the evolution of the constraints is given by
\begin{equation}
 \phi^\epsilon_j(n+1)=\phi^\epsilon_j(n)+C_{jk}^{\epsilon \enspace m}
 \phi_m^\epsilon \lambda_k^\epsilon \delta
 +A^\epsilon_{jk}\lambda_k^\epsilon \delta +\mathcal{O}(\delta^2)\,.
\end{equation}
Taking the limits $\delta\rightarrow 0$ and \(\epsilon\rightarrow0\)
we recover the evolution equations for the constraints in the
continuum theory
\begin{equation}
 \dot{\phi}_j=\lim_{\epsilon,\delta\rightarrow 0}
 \frac{\phi^\epsilon_j(n+1)-\phi^\epsilon_j(n)}{\delta}
 =C_{jk}^{\enspace \enspace m}\phi_m \lambda_k\,,
\end{equation}
where $\lambda_i=\lim_{\epsilon \rightarrow 0}\lambda_i^\epsilon$ are Lagrange
multipliers. Let us remark that although at first sight the
$\lambda_i^\epsilon$ seem not to be free in the discrete theory,
they are only determined by the constraints evaluated on the initial
data, and thus can be chosen arbitrarily.

Applying the Dirac procedure to the discrete theory would lead us to
different degrees of freedom in the continuum theory, due to the
fact that the constraints in this case are second class. This can be
shown in the following way. Within the Dirac procedure one would
impose the constraints \(\phi_i^\epsilon\) and construct the total
Hamiltonian \(H_T=C^j\phi_j^\epsilon\) with a number $M$ of Lagrange
multipliers \(C_j\). Consistency together with the discrete
constraint algebra (\ref{eq:dca}) would require
\begin{equation}
 C^jA_{jk}^\epsilon=0\,.
\end{equation}
Therefore the \(C_j\) would generally not be free but restricted.
This implies that one would have more observables than in the
continuum theory, i.e.,more than \(2N-2M\), so the degrees of
freedom of the discretized theory would not match the ones of the
continuum theory.  Thus, the uniform discretization method is better
suited for studying a discretized, totally constrained system.

The uniform discretization approach provides a method to handle the
fact that the constraints become second class through
discretization, without changing the number of degrees of freedom.
Moreover the discrete constraint algebra mimics the one of the
continuum.

\section{Spherically symmetric gravity coupled to a scalar field}
\label{sec:Spher}

In this section we review the treatment of the spherically symmetric
gravity in Ashtekar variables. This symmetry reduction was first
performed by Bengtsson \cite{Bengtsson90}. Later on, the definition
of the invariant connection components was slightly changed to
ensure correct transformation properties \cite{Thiemann93}. We will
use the invariant connection formulation based on the deep
investigation on symmetry reduction of connections in
\cite{Bojowald00}. Using the invariant formulation of the basic
variables we find the constraints in the spherical symmetric
framework. Furthermore we will couple a spherical symmetric massless
scalar field to gravity, following \cite{Gambini09}. Subsequently we
perform a gauge fixation in order to solve the diffeomorphism
constraint. Since the spherical symmetric theory is manifestly gauge
invariant we are left with the Hamiltonian constraint.  The
variational method we are using is based on the uniform
discretization method and thus we end this section with the
discretization of the Hamilton constraint.

In order to describe spherically symmetric spacetimes we use
spherical coordinates $(r,\varphi,\theta)$ on a spatial manifold
with topology $\Sigma=\R^+\times S^2$. The invariant connection,
densitized triad and extrinsic curvature can be written as:
\begin{eqnarray}
A&=&A_r\Lambda_3 dr+\left(A_1\Lambda_1+A_2\Lambda_2\right)d\theta
+\left(\left(A_1\Lambda_2-A_2\Lambda_1\right)\sin\theta
+\Lambda_3 \cos\theta\right) d\varphi\,,\\
E&=&E^r\sin \theta \Lambda_3 \, {\partial_r}+
\left(E^1\Lambda_1+E^2\Lambda_2\right)\sin \theta
\,{\partial_\theta}+
\left(E^1\Lambda_2-E^2\Lambda_1\right)\,{\partial_\varphi}\,,\\
K&=&K_r\Lambda_3 \,
dr+\left(K_1\Lambda_1+K_2\Lambda_2\right)\,d\theta+
\left(K_1\Lambda_2-K_2\Lambda_1\right)\sin\theta \,d\varphi\,,
\end{eqnarray}
where $A_i,E^i,K_i$ ($i=r,1,2$) are arbitrary functions on $\R^+$
and $\Lambda_I$ are generators\footnotemark\, of $\su(2)$\,
\cite{Bojowald06,Bojowald00}.
\footnotetext{$\Lambda_I=-i\sigma_I/2$, where $\sigma_I$ denotes a
rigid rotation of the Pauli matrices.}

In order to get canonically conjugate variables and suitable fall
off conditions we use the extended ADM phase space variables
\cite{Bojowald:2004af,Bojowald06}. For convenience we introduce the
following change of variables
\begin{align}
K_\varphi(r)=\sqrt{K_1(r)^2+K_2(r)^2}\,,\qquad
E^\varphi(r)=\sqrt{E^1(r)^2+E^2(r)^2}\,,
\end{align}
that implies the Poisson relations
\begin{align}
 \left\{K_r(x),E^r(x')\right\}&=2 G \delta(x-x')\,,\\
 \left\{K_\varphi(x),E^\varphi(x')\right\}&=G \delta(x-x').
 \label{eq:PEK}
\end{align}
In terms of the canonical conjugated variables
\(\{K_r,E^r,K_\varphi,E^\varphi\}\) we can write the constraints as:
\begin{eqnarray}
\mathcal{G}&=&0\,,\\
G\, C&=&K_\varphi'E^\varphi-\frac{1}{2} K_r (E^r)'
\label{eq:sphercon1}\,,\\
G\, H&=&\frac{\sqrt{E^r}(E^r)''}{2 E^\varphi}
-\frac{\sqrt{E^r}(E^r)'(E^\varphi)'}{2 (E^\varphi)^2} -\sqrt{E^r}K_r
 K_\varphi+\frac{((E^r)')^2}{8 E^\varphi\sqrt{E^r}}
 -\frac{E^\varphi K_\varphi^2}{2 \sqrt{E^r}}
 -\frac{E^\varphi}{2 \sqrt{E^r}}\,,\label{eq:sphercon2}
\end{eqnarray}
where $\mathcal{G}$, $C$ and $H$ stands for the Gau\ss,
diffeomorphism and Hamiltonian constraint respectively; $G$ is the
Newton constant; and $f'$ denotes the derivative of the function $f$
with respect to the radial coordinate. Notice that the variables are
manifestly gauge invariant.

Now we add the constraints corresponding to the matter content (the
massless scalar field $\phi$) \cite{Husain05}:
\begin{eqnarray}
 C_\text{matt}&=&P^{\phi}\phi', \label{eq:mattcon1}\\
{\tilde{H}}_\text{matt}&=&\frac{1}{2 E^\varphi
\sqrt{|E^r|}}\left((E^r)^2 (\phi')^2+(P^\phi)^2\right)\,.
\label{eq:mattcon2}
\end{eqnarray}

Then, the Hamiltonian and diffeomorphism constraint of the reduced
theory minimally coupled to the massless scalar field are
\begin{eqnarray}
H&=&\frac{1}{G}\left[-\frac{E^{\varphi}}{2\sqrt{|E^r|}}- K_\varphi
\sqrt{|E^r|}K_r-\frac{E^\varphi K_\varphi^2}{2\sqrt{|E^r|}}+
\frac{(|E^r|')^2}{8\sqrt{|E^r|}E^{\varphi}}
-\frac{\sqrt{|E^r|}|E^r|'(E^\varphi)'}{2(E^\varphi)^2}
+\frac{\sqrt{|E^r|}(|E^r|)''}{2E^\varphi}\right]\nn\\
&& +\frac{(P^\phi)^2}{2\sqrt{|E^r|}E^\varphi}
+\frac{|E^r|^{3/2}(\phi')^2}{2E^\varphi}\,,
\\
C&=&\frac{1}{G}\left(E^\varphi K_\varphi' -\frac{1}{2}(E^r)'K_r
\right) + P_\phi \phi'\,.
\end{eqnarray}

Now, we proceed by gauge fixing \(E^r=r^2=(x+a)^2\)\, as in
\cite{Campiglia07}, where $x$ is a radial coordinate and $a$ is a
constant that we will relate  with the Schwarzschild radius later
on. Then we solve the diffeomorphism constraint, which leads to
\begin{align}
K_r=\frac{E^\varphi(K_\varphi)'}{r}+G \frac{P_\phi \phi'}{r}\,.
\end{align}
Rescaling the lapse function \(N\rightarrow N G 2r/E^\varphi\) one
gets for the Hamiltonian constraint
\begin{eqnarray}
H&=&H_\text{vac}+ G\, H_\text{matt}\,, \label{eq:Hamilton}\\
H_\text{vac}&=&\left(-x
-(x+a) K^2_\varphi
+\frac{(x+a)^3}{(E^\varphi)^2}\right)'\,,\\
H_\text{matt}&=&\frac{P_\phi^2}{(E^\varphi)^2}
+\frac{(x+a)^4(\phi')^2}{(E^\varphi)^2}-2\frac{(x+a) K_\varphi
P_\phi \,\phi'}{E^\varphi}.
\end{eqnarray}

\bigskip

Now, we proceed to discretize and polymerize the Hamiltonian
constraint. We consider
\begin{eqnarray*}
&& r  \rightarrow  r(i)\,,\qquad H  \rightarrow
\frac{H(i)}{\epsilon}\,,\qquad \phi(r)  \rightarrow \phi(i)\,,\qquad
P^\phi
\rightarrow  \frac{P^\phi(i)}{\epsilon}\,,\\
&& E^\varphi \rightarrow \frac{E^\varphi(i)}{\epsilon}\,,\qquad
K_\varphi  \rightarrow \frac{\sin(\rho
K_{\varphi}(i))}{\rho}\,,\qquad f' \rightarrow
\frac{f(i+1)-f(i)}{\epsilon}\,.
\end{eqnarray*}
For the sake of simplicity, we will use a constant parameter
\(\rho\) for the polymerization, like in the early loop quantum
cosmology (LQC).

Finally, the expression for the discretized Hamiltonian constraint
is
\begin{align}
\begin{split}
H(i)=&-(1-2\Lambda)\epsilon+r(i+1)\frac{\sin^2(\rho K_{\varphi}(i+1))}{\rho^2}
+r(i)\frac{\sin^2(\rho
K_{\varphi}(i))}{\rho^2}+\frac{r(i+1)^3\epsilon^2}{E^\varphi(i+1)^2}
-\frac{r(i)^3\epsilon^2}{E^\varphi(i)^2}\\
&+\ell_p^2\left[\epsilon\frac{P^\phi(i)^2}{E^\varphi(i)^2}
+\epsilon\frac{r(i)^4}{E^\varphi(i)^2}\left(\phi(i+1)-\phi(i)\right)^2
-2\,\frac{r(i)P^\phi(i)\sin(\rho K_{\varphi}(i))}{E^\varphi(i) \rho
} \left(\phi(i+1)-\phi(i)\right)-\epsilon \rho_\text{vac}\right],
\end{split}
\label{eq:disHam}
\end{align}
where we are employing units such that $G=\ell_p^2,\, c=\hbar=1$. We
have also introduced a cosmological constant \(\Lambda\) which is
related to the vacuum energy of the scalar field
\(\rho_\text{vac}=\frac{2}{G}\Lambda\). We need to introduce this
term in order to consider the Fock vacuum of the scalar field. This
vacuum energy term comes from the fact that we impose the
expectation value of the matter constraint in the semiclassical
limit $\ell_p\rightarrow 0$ to vanish. The value of
\(\rho_\text{vac}\) can then be evaluated by introducing the Fock
vacuum in a fixed background setting (we will choose our vacuum to
be the one defined in the asymptotic region of the Schwarzschild
spacetime) and computing the ``energy of the vacuum'', i.e., the
expectation value of the matter Hamiltonian.  We  find that the
vacuum energy at the order of interest is
\[\rho_\text{vac}=\frac{\pi}{2\epsilon^2}\,.\]

As pointed out in \cite{Gambini09}, the introduction of the
cosmological constant has some disadvantages. Introducing a
cosmological constant before the symmetry reduction implies ending
up with a non constant term in two dimensions, which cannot give
rise to the vacuum energy. Because of this, the theory does not stem
from a dimensional reduction of the full theory if one introduces
the cosmological constant in the way we did. To circumvent this
problem one can use the fact that, unlike in the full four
dimensional theory, there is already a constant term in the
Hamiltonian constraint. The cosmological constant at that level can
be interpreted as a rescaling of the radial coordinate. However this
rescaling has the disadvantage that the volume of spheres is not
\(4\pi R^2\) anymore, i.e., the full theory one approximates has
topological defects.

\section{Construction of quantum trial states}
\label{sec:gravtrial}

Although the discrete Hamiltonian constraint (\ref{eq:disHam}) fails
to close a first class algebra, it can be shown that using the
uniform discretization approach one can consistently treat the
problem by minimizing the resulting Hamiltonian constraint. In order
to achieve that, we use the variational technique described in
\cite{Gambini09}.

We have the opportunity to consider the gravitational sector and the
matter sector separately, since classically the scalar field
vanishes in the vacuum. We thus construct a polymeric Hilbert space
$\mathcal{H}_{\text{poly}}$ for the gravitational variables. The
gravitational trial state  $|\Psi_{\vec{\sigma}}\rangle$ is then
constructed as a Gaussian in phase space with width $\sigma$
centered around the Schwarzschild solution. With the help of this
trial state we find an effective Hamiltonian for the matter
variables. Out of the equations of motion for this Hamiltonian we
construct solutions based on creation and annihilation operators.
Thus we introduce the Fock vacuum state $|0\rangle$ for the matter
sector. In this way, we make contact with the usual treatments of
quantum field theory in curved spacetimes.

The trial state used for the minimization of the master constraint
is then constructed as a direct product of the gravitational
Gaussian state and the Fock vacuum
\begin{align}
|\Psi_{\vec{\sigma}}^\text{trial}\rangle=|\Psi_{\vec{\sigma}}\rangle
\otimes |0\rangle\,. 
\end{align}

\subsection{Gravitational part of the trial states}

In the usual LQC treatment one uses cylindrical functions, which are
related with the connection variables through the holonomies (see
e.g. \cite{Thiemann,Bojowald:2008zzb}). Here, we mimic this method
for the spherical symmetric theory. First, we set up a polymeric
Hilbert space $\mathcal{H}_\text{poly}$ with a spin network like
basis and define the action of the operators $E^\varphi$ and
$K_\varphi$ on this state. We will then proceed by defining the
gravitational part of our trial states centered around the classical
Schwarzschild solution.

In order to use the uniform discretization method, we consider a
discretized setting. The symmetry reduced spatial manifold in the
spherically symmetric case is a dimensional line (the radial
direction). In analogy with LQC, we consider the bulk Hilbert space
for the gravitational sector as
\begin{align}
 \mathcal{H}_\text{poly}=L^2(\otimes_N \bar{R}_\text{Bohr},\otimes_N
 d\mu_0)\,,
\end{align}
where $\bar{R}_\text{Bohr}$ is the Bohr compactification of the real
line, $d\mu_0$ is the Haar measure and $N$ is the number of cells in
the discretization. In this framework, the basis for a fixed graph
$g$ composed by $N$ edges \(e_j\) and vertices $v_j$ (with
$j=1,\ldots,N$) is:
\begin{align}
\langle K_\varphi(j)|\vec{\mu}\rangle=\prod_{j} \exp(i\mu_j
K_\phi(j))\,,
\end{align}
with \(\mu_j \in \mathbb{R}\). The variables satisfy the classical
Poisson bracket
\begin{align}
 \{K_\varphi(i),E^\varphi(j)\}=G\delta_{ij}\,,
\end{align}
which suggests defining \(E^\varphi\) as a derivation operator
\begin{align}
 \hat{E}^\varphi(i)=-\ell_p^2\frac{\partial}{\partial K_\varphi(i)}\,.
\end{align}
Its action over the elements \( | \vec{\mu} \rangle\) of the basis
of the gravitational Hilbert space is defined by
\begin{align}
 \hat{E}^{\varphi}(i) | \vec{\mu} \rangle =
 \ell_p^2 \mu_i | \vec{\mu}\rangle\,.
\end{align}
We associate the \(K_\varphi\) to point holonomies, and their action
is defined by
\begin{align}
\exp(i\rho \hat{K}_{\varphi}(i))| \vec{\mu} \rangle=| \vec{\mu}+\rho
\vec{e}_i\rangle. \label{eq:Kev}
\end{align}
We want to remark that the Hilbert space we are using is a direct
product of the Hilbert space of LQC for every lattice position
\(i\), which enables us to use techniques developed in LQC for our
purpose.

Finally, we construct the trial states as
\begin{equation}
|\Psi_{\vec{\sigma}}\rangle=\langle \vec{\mu} \mid
\psi_{\vec{\sigma}} \rangle=\prod_{i}{\sqrt[4]{\frac{2}{\pi
\sigma(i)}}\exp\left(-\frac{1}{\sigma(i)}
\left(\mu_i-\frac{r_1(i)\epsilon}{\ell_p^2}\right)^2\right)}\,,
\label{eq:trial}
\end{equation}
with $\mu_i$ centered at
$E^\varphi(i)=\ell_p^2\mu_i=\epsilon\,r_1(i)$, where the value of
$r_1(i)$ will be determined in the following.

Since the classical counterpart of the vacuum solution corresponds
to vanishing scalar fields, we can ignore the matter part of the
Hamiltonian $H_\textrm{matt}$ and focus on the gravitational part
$H_\textrm{vac}$. Demanding \(H_\text{vac}=0\) and additionally
using the gauge \(K_\varphi=0\) one gets
\begin{equation*}
 \left(-x(1-2\Lambda)+\frac{(x+a)^3}{(E^\varphi)^2}\right)'=0\quad
 \Rightarrow \quad E^\varphi=\frac{1}{\sqrt{1-2\Lambda}}\frac{x+a}{\sqrt{1-\frac{a-C}{x+a}}}\,,
 \label{eq:ephi}
\end{equation*}
where $C$ is an integration constant. In order to recover the
Schwarzschild metric for $\Lambda=0$, we set $C=0$. Then we obtain
\begin{align}
r_1(i)=\frac{1}{\sqrt{1-2\Lambda}}\frac{r(i)}{\sqrt{1-\frac{a}{r(i)}}}\,.
\end{align}
We find that, in contrast to \cite{Campiglia07}, \(a\) is not the
Schwarzschild radius \(R_S\) but rather
 \[a=R_S\sqrt{1-2\Lambda}.\]
 As it was commented before, the introduction of the cosmological
 constant induces a rescaling of the radial coordinate.

\subsection{Matter part of the trial states}

In order to construct the matter part of the trial states we define
an effective Hamiltonian as the expectation value of the matter
Hamiltonian constraint over the gravitational part of the trial
states (\ref{eq:trial}). We compute the equations of motion
corresponding to this effective Hamiltonian and get a differential
equation for the matter variables \(\phi\) and \(P^\phi\). Working
in the Fourier space, this equation turns out to be a
Sturm-Liouville problem. We can construct the associated creation
and annihilation operators without knowing the explicit solution to
the equation (making use of the tools of the Sturm-Liouville
theory). It is thus straightforward to introduce the Fock-vacuum
state, which completes the construction of the trial state.

Due to the usual factor ordering ambiguities in the quantization
procedure, we need to choose one prescription for it. Here, we use
the factor ordering of \cite{Gambini09}, namely putting the
operators \(\hat{E}^\varphi\) and \(\hat{P}^\phi\)  symmetrically
around \(\hat{K}_\varphi\) and \(\hat{\phi}\) respectively. In this
way, the discretized matter Hamilton constraint operator can be
written as:
\begin{align}
\begin{split}
\hat{H}_\text{matt}(i)=&\epsilon\left(
\hat{P}^\phi(i)^2+r(i)^4(\hat{\phi}(i+1)-\hat{\phi}(i))^2\right)
\frac{1}{(\hat{E}^\varphi(i))^2}
\\
&- 2\frac{r(i)}{\rho}\sqrt{\hat{P}^\phi}
(\hat{\phi}(i+1)-\hat{\phi}(i)) \sqrt{\hat{P}^\phi}
\frac{1}{\sqrt{\hat{E}^\varphi}}\sin(\rho\hat{K}_{\varphi}(i)
)\frac{1}{\sqrt{\hat{E}^\varphi}}-\epsilon\, \rho_\text{vac}\,.
\end{split}
\label{eq:dismattHam}
\end{align}
Using the expectation values (\ref{eq:Sin/E}) and (\ref{eq:1/E2}) we
obtain the effective Hamiltonian constraint
\begin{eqnarray}
&&\hat{H}_\text{matt}^\text{eff}=\langle\Psi_{\vec{\sigma}}|
\hat{H}_{\text{matt}}|\Psi_{\vec{\sigma}}\rangle
=\left(\left(\hat{P}^\phi\right)^2+r^4\left(\hat{\phi}'\right)^2\right)
f(r)-\rho_\text{vac}\,,\label{eq:effHam}
\\
&&f(r)=\frac{1}{r_1^2}+\frac{1}{r_1^4}\,
\frac{\ell_p^4}{\epsilon^2}\alpha
+\mathcal{O}\left(\frac{\ell_p^8}{\epsilon^4}\right),\nn
\end{eqnarray}
where we went back to the continuum theory because it is easier to
solve the corresponding differential equations than the original
difference equations.

The equations of motion for the effective Hamiltonian constraint
(\ref{eq:effHam}) are:
\begin{align}
\dot{\phi}(r,t)&=\frac{\delta H_\text{matt}^\text{eff}}
{\delta P^\phi}=2 f(r) P^\phi(r,t)\,, \label{eq:Hameq1}
\\
\dot{P^\phi}(r,t)&=-\frac{\delta H_\text{matt}^\text{eff}}{\delta
\phi}=\left(2 f(r)r^4\phi'(r,t)\right)'\,.
\end{align}
Now, we consider the Fourier like transformation
\begin{eqnarray}
&&\phi(r,t)=\int_0^\infty \frac{\phi(r,\omega)}{\sqrt{ 2
\omega}}\left(e^{i\omega t} \bar{C}(\omega)+e^{-i\omega t}
C(\omega)\right)d\omega\,, \label{eq:phi_clas}\\
&&P^{\phi}(r,t)=\frac{1}{2 f(r)}\frac{\partial \phi(r,t)}{\partial
t}=i A(r) \int_0^\infty \sqrt{\frac{\omega}{2}}\,
\phi(r,\omega)\left(e^{i\omega t} \bar{C}(\omega)-e^{-i\omega t}
C(\omega)\right)d\omega\,,\label{eq:Pphi_clas}
\end{eqnarray}
where the transformation for the momentum follows from
(\ref{eq:Hameq1}). This leads to the following Sturm-Liouville
differential equation
\begin{equation}
\left(B(r) \phi'(r,\omega)\right)'+\omega^2 A(r) \phi(r,\omega)=0\,,
\label{eq:SL}
\end{equation}
with
\begin{align}
A(r)=\frac{1}{2 f(r)}\,,\qquad\, B(r)=2 f(r)r^4.
\end{align}
According to the Sturm-Liouville theory, solutions of such
differential equations form a basis of the Hilbert space
\(L^2(\mathbb{R},A(r)dr)\). This means that properly normalized
functions \(\phi(r,\omega)\) satisfy the orthogonality and closure
relations
\begin{eqnarray}
&&\int_0^\infty{dr\,A(r)\,\phi(r,\omega)\phi(r,\omega')
=\delta\left(\omega-\omega'\right)}\,, \label{eq:ortho}
\\
&&\int_0^\infty{d\omega\,\phi(r,\omega)\phi(r',\omega)
=\frac{1}{A(r)}\delta\left(r-r'\right)}\,.
\label{eq:closure}
\end{eqnarray}
Consequently, the field and momentum operators are constructed as:
\begin{eqnarray}
&&\hat{\phi}(r,t)=\int_0^\infty \frac{\phi(r,\omega)}{\sqrt{ 2
\omega}}\left(e^{i\omega t} \hat{C}^\dagger(\omega)+e^{-i\omega t}
\hat{C}(\omega)\right)d\omega\,, \label{eq:phi}\\
&&\hat{P}^{\phi}(r,t)=i A(r) \int_0^\infty \sqrt{\frac{\omega}{2}}\,
\phi(r,\omega)\left(e^{i\omega t}
\hat{C}^\dagger(\omega)-e^{-i\omega t}
\hat{C}(\omega)\right)d\omega\,.\label{eq:Pphi}
\end{eqnarray}

Considering the commutation relation
$\left[\hat{C}(\omega),\hat{C}^\dagger(\omega')\right]
=\delta(\omega-\omega')$\, and making use of the closure relation,
we obtain the standard commutator for the field and its conjugate
momentum operators
\[\left[\hat{\phi}(r,t),\hat{P}^{\phi}(r',t)\right]
=i\delta\left(r-r'\right)\,.\]
As it is inferred by the previous commutation relations, the
operators \(\hat{C}\) and \(\hat{C}^\dagger\) act as annihilation
and creation operators respectively.

Finally, we construct the trial state as the direct product of the
Gaussian state (\ref{eq:trial}) and the Fock vacuum of the matter
part:
\begin{align}
|\Psi_{\vec{\sigma}}^\text{trial}\rangle=|\Psi_{\vec{\sigma}}\rangle
\otimes |0\rangle\,, \label{eq:full_trial_state}
\end{align}
where $|0\rangle$ is the state annihilated by the annihilation
operator \(\hat{C}\). In this treatment, by construction, the case
of entanglement between matter and gravitational variables is
excluded,  because the split between the gravitational and the
matter sector works only with the vacuum state. For excited states
this requirement should be relaxed.

\section{Expectation value of the master constraint}
\label{sec:expectation}

In this section we use the trial state (\ref{eq:full_trial_state})
to calculate the expectation value of the master constraint. First,
we construct the discrete master constraint out of
(\ref{eq:disHam}). We are able to write the principal order of the
expectation value in orders of the lattice spacing. It shows that
the master constraint is becoming small for bigger lattice spacings.
Since the approximations made break down at a certain point one
concludes that there is a minimum of the master constraint. However,
the minimum value is not reached in the limit $\epsilon\rightarrow
0$. This fact suggests that there is no continuum limit.

\subsection{The discrete master constraint}

Using the uniform discretization technique we construct the master
constraint (\ref{eq:unimas}) as
\begin{equation}
\mathbb{H}=\frac{1}{2}\int{dr\frac{H(r)^2}{r E^\varphi}\,\ell_p}\,,
\end{equation}
where we used our gauge and took into account that the Hamiltonian
is a density of weight one, so we added the determinant of the
3-metric in order to make contact with the continuum theory. The
Planck length $\ell_p$ ensures that $\mathbb{H}$ is dimensionless.
Notice that the master constraint is proportional to $1/E^\varphi$.
This implies that the vacuum state corresponds to the zero loop
state ($\rho=0$) which is degenerate in the polymer representation.

Nevertheless, the Hamiltonian constraint can be rescaled by a scalar
function of the canonical variables without changing the first class
nature of the constraint algebra, so we can solve this problem by
rescaling $H(x)\rightarrow H(x)\sqrt{E^\varphi/(E^x)'}$. Then, in
the discrete theory we obtain
\begin{equation}
\mathbb{H}^\epsilon=\sum_i{\mathbb{H}(i)}=
\sum_i{\frac{1}{4}\frac{H(i)^2}{\epsilon r(i)^2}\ell_p}\,.
\end{equation}
For computational purposes, we proceed to separate ``matter'' and
``gravity'' operators acting on their corresponding Hilbert spaces,
i.e., we write the Hamiltonian constraint (\ref{eq:disHam}) as
\begin{equation}
H(i)=H^{(1)}_{\text{matt}}(i) c_1(i)+H^{(2)}_{\text{matt}}(i) c_2(i)+H^{(3)}_{\text{matt}}(i) c_3(i)+c_4(i)\,,
\end{equation}
where
\begin{align}
H^{(1)}_{\text{matt}}(i)&=\ell_p^2\left(\epsilon P^\phi(i)^2
+\epsilon r(i)^4(\phi(i+1)-\phi(i))^2\right) \label{eq:H1}\,,\\
H^{(2)}_{\text{matt}}(i)&=\ell_p^2\left(-2r(i)P^\phi (\phi(i+1)
-\phi(i))\right)\label{eq:H2}\,,\\
H^{(3)}_{\text{matt}}(i)&=l_p^2\,\rho_\text{vac}\,,
\end{align}
and
\begin{align}
\begin{split}
c_1(i)=&\frac{1}{E^\varphi(i)^2}\,,\qquad
c_2(i)=\frac{\sin(\rho K_{\varphi}(i))}{E^\varphi(i) \rho}\,,\qquad
c_3(i)=-1\,,\\
c_4(i)=&-(1-2\Lambda)\epsilon-\left(r(i+1)\frac{\sin^2(\rho
K_{\varphi}(i+1))}{\rho^2}-r(i)\frac{\sin^2(\rho
K_{\varphi}(i))}{\rho^2}\right)
+\frac{r(i+1)^3\epsilon^2}{E^\varphi(i+1)^2}
-\frac{r(i)^3\epsilon^2}{E^\varphi(i)^2}\,.
\end{split}
\end{align}
In terms of these operators, we can write the expression for the
discretized master constraint in the following form:
\begin{align}
\mathbb{H}(i)=\ell_p\frac{1}{4 \epsilon r(i)^2}&\left( c_{11}(i)(
H^{(1)}_{\text{matt}}(i))^2+c_{22}(i) ( H^{(2)}_{\text{matt}}(i))^2+c_{33}(i)( H^{(3)}_{\text{matt}}(i))^2\right.\nn\\
&+2 c_{12}(i)
H^{(1)}_{\text{matt}}(i)H^{(2)}_{\text{matt}}(i)
+2c_{13}(i)H^{(1)}_{\text{matt}}(i)H^{(3)}_{\text{matt}}(i)+2 c_{23}(i) H^{(2)}_{\text{matt}}(i)H^{(3)}_{\text{matt}}(i)\\
&\left.+2 c_{14}(i) H^{(1)}_{\text{matt}}(i)+2 c_{24}(i) H^{(2)}_{\text{matt}}(i)+2 c_{34}(i) H^{(3)}_{\text{matt}}(i)
+c_{44}(i) \right) \,, \label{eq:dismas}\nn
\end{align}
where $c_{jk}(i)=c_j(i)\cdot c_k(i)$.

As the matter and the gravitational sector are not entangled, i.e.,
$\langle0|\hat{c}_{jk}(i)|0\rangle=\hat{c}_{jk}(i)$ and
$\langle\Psi_{\vec{\sigma}}|\hat{H}^{(j)}_{\text{matt}}(i)|
\Psi_{\vec{\sigma}}\rangle=\hat{H}^{(j)}_{\text{matt}}(i)$, we can
accomplish the computation of the expectation values separately.

\subsection{Minimizing the master constraint}

In this part we treat the problem of minimizing the master
constraint. All the intermediate calculations are included in the
appendices. For the sake of simplicity, we go back to the continuum
theory and assume that \(\sigma\) is independent of the lattice
position, which means:
\begin{align}
\begin{split}
&r(i)\rightarrow r\,,\qquad
r(i+1)\rightarrow r+\epsilon\,,\\
&\sigma(i)\rightarrow \sigma\,,\qquad \sigma(i+1) \rightarrow
\sigma\,.
\end{split}
\end{align}
Because of the independence of \(\sigma\) with respect to the
position we also set \(\sigma=\sigma_0\frac{\epsilon^n}{\ell_p^n}\).

As it was explained in \cite{Gambini09}, the approximation we
considered in order to handle the expressions (neglecting higher
powers of $\epsilon$) is inadequate for large values of $\epsilon$.
More specifically, the approximation is valid up to values of
\(\epsilon\approx10^{-23}\,\text{cm} \). In this range it is
convenient to sort the terms in orders of
\(\mathcal{O}((\ell_p/\epsilon)^k\epsilon)\). Then, the non
vanishing components of the master constraint (\ref{eq:dismas}) up
to the principal order are:
\begin{align}
&\langle c_{11}( H^{(1)}_{\text{matt}})^2
\rangle=\left(3A(r)^4I_1^2+3 r^8 I_2^2+2A(r)r^4\left(2I_3^2+I_1
I_2\right)\right)\left[\frac{\ell_p^4 \epsilon^2}{4
r_1^2}+\mathcal{O}\left(\frac{\ell_p^k}{\epsilon^k}\epsilon^8
\right)\right]\,,\\
&\langle c_{13} H^{(1)}_{\text{matt}} H^{(3)}_{\text{matt}}  \rangle=\left(A(r)^2I_1^2+ r^4 I_2\right)\left[-\frac{l_p^2 \epsilon^2}{ r_1^2}\Lambda+\mathcal{O}\left(\frac{l_p^k}{\epsilon^k}\epsilon^6 \right)\right]\,,\\
&\langle c_{14} H^{(1)}_{\text{matt}}  \rangle=\left(A(r)^2I_1^2+
r^4 I_2\right)\left[\left(\frac{1}{2\rho^2}\frac{1}{r_1^4}
\left(1-\exp\left(-\frac{2\rho^2}{\sigma_0}\frac{\ell_p^n}{\epsilon^n}
\right)\right)+\frac{3r^2}{r_1^4}\right)\ell_p^2\epsilon^2+
\mathcal{O}\left(\frac{\ell_p^k}{\epsilon^k}\epsilon^6
\right)\right]\,,
\\
&\langle c_{22}( H^{(2)}_{\text{matt}})^2 \rangle= r^2
A(r)^2\left(2I_3^2+ I_1
I_2\right)\left[\frac{1}{2\rho^2}\left(1-\exp\left(
-\frac{2\rho^2}{\sigma_0}\frac{\ell_p^n}{\epsilon^n}\right)\right)
\frac{\ell_p^2\epsilon^2}{r_1^2}+\mathcal{O}\left(
\frac{\ell_p^k}{\epsilon^k}\epsilon^8\right)\right]\,,
\\
&\langle c_{33}( H^{(3)}_{\text{matt}})^2 \rangle=4\Lambda^2\epsilon^2\,,
\\
&\langle c_{34} H^{(3)}_{\text{matt}} H^{(4)}_{\text{matt}}  \rangle=\frac{1}{2\rho^2}\left(1-\exp\left(-\frac{2\rho^2}{\sigma_0}\frac{l_p^n}{\epsilon^n}\right)\right)2\Lambda^2\epsilon^2+\mathcal{O}\left(\frac{l_p^k}{\epsilon^k}\epsilon^3 \right)\,,
\\
&\langle c_{44}
\rangle=\frac{r^2}{4\rho^4}\left(1-2\exp\left(-\frac{4\rho^2}{\sigma_0}
\frac{\ell_p^n}{\epsilon^n}\right)+\exp\left(-\frac{8\rho^2}{\sigma_0}
\frac{\ell_p^n}{\epsilon^n}\right)\right)+\mathcal{O}\left(
\frac{\ell_p^k}{\epsilon^k}\epsilon\right)\,,
\end{align}
where $I_1, I_2$ and $I_3$ are integrals coming from the computation
of the expectation value of the Hamiltonian matter constraint and
are given by
\begin{align}
\label{eq:I1_1} I_1=\int_0^\infty d\omega  \, \omega
(\phi(r,\omega))^2\,,\qquad I_2=\int_0^\infty d\omega \frac{1}{
\omega } (\phi'(r,\omega))^2\,,\qquad I_3=\int_0^\infty d\omega
\,\phi(r,\omega)\phi'(r,\omega)\,,
\end{align}
where $\phi(r,\omega)$ is the solution of the Sturm-Liouville
equation (\ref{eq:SL}). In the appendix \ref{app:coeff} we give the
expressions (\ref{eq:I11}) of these integrals for an approximate
solution $\tilde{\phi}_0(r,\omega)$ of the zeroth order of the
Sturm-Liouville equation (\ref{eq:asol}).

Finally, the main order of the expectation value of the integrand of the
master constraint takes the form
\begin{align}
\langle \mathbb{H}(r) \rangle=
&\frac{\ell_p}{\epsilon}\frac{1}{16\rho^4}\left(1-2\exp\left(
-\frac{4\rho^2}{\sigma_0}\frac{\ell_p^n}{\epsilon^n}\right)
+\exp\left(-\frac{8\rho^2}{\sigma_0}\frac{\ell_p^n}{\epsilon^n}\right)
\right)+\mathcal{O}\left(\frac{\ell_p^k}{\epsilon^k}\epsilon\right)\,.
\end{align}
Notice that the leading order term does not depend on the variable
$a$. Also, since the equations are lengthy and they do not provide
any further conclusions, we leave away higher order corrections.

\begin{figure}[htbp]
    \centering
        \includegraphics[width=0.70\textwidth]{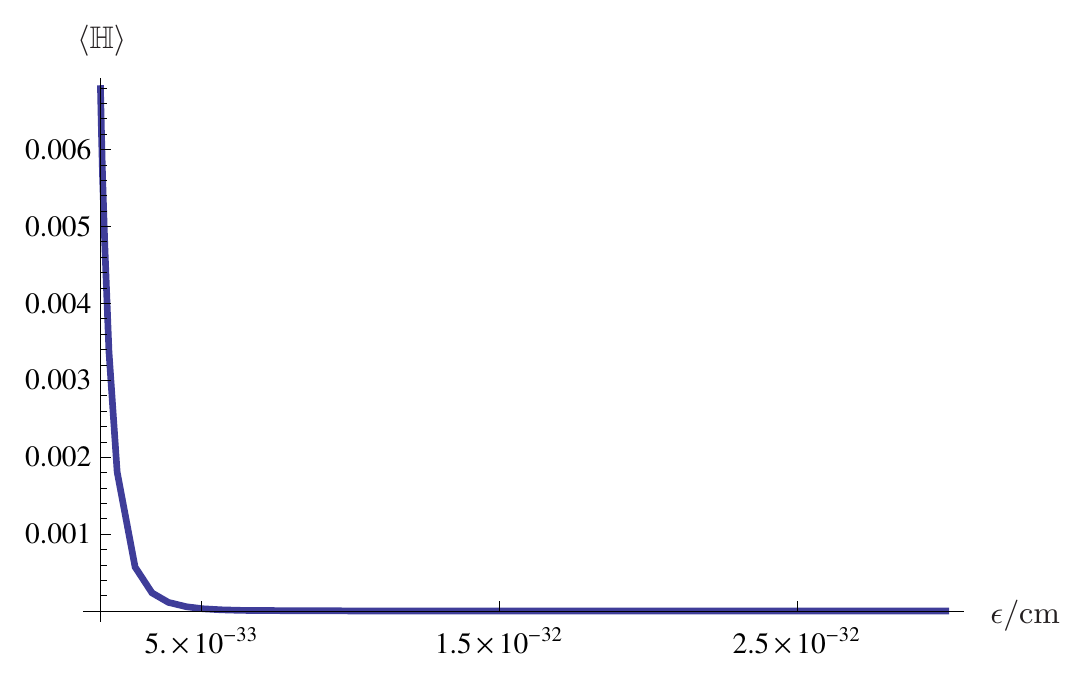}
        \caption{Expectation value of the integrand of the master
        constraint for \(\sigma_0=10, n=2\) and \(\rho=1\)
    as a function of the lattice spacing \(\epsilon\) in the region
    near \(\ell_p\).}
    \label{figmas}
\end{figure}

We are now in conditions to study the minimum of this master
constraint. In figure \ref{figmas} we plot the expectation value
(for values \(\sigma_0=10, n=2\) ) with respect to the lattice
spacing $\epsilon$ in the region of interest. We observe that the
master constraint drops very fast for lattice spacings larger than
the Planck scale. As in the case studied in \cite{Gambini09},
because of the break down of the approximation for
$\epsilon>10^{-23}\text{cm}$, we expect the master constraint to
increase again. So, we can conclude that there is a minimum around
$\epsilon\approx 10^{-23}\,\text{cm}$.

We find that varying the variable $a$, which is connected to the Schwarzschild
radius of the classical solution where the trial states are
centered, has influence on the master constraint only at the
subleading orders. So we can conclude that, as in the Minkowski case
studied in \cite{Gambini09}, we can construct the vacuum state for
our minimally coupled system.

\section{Conclusions}

In this paper we worked out an approximation for the vacuum state of
a scalar field coupled to gravity with spherical symmetry. We
focused on the Schwarzschild spacetime for the gravitational sector
and we employed the method presented in \cite{Gambini09}.

The vacuum for the coupled system is given by the direct product of
the Fock vacuum state for the scalar field and a Gaussian centered
around the classical Schwarzschild solution for the gravitational
sector. In order to deal with the dynamics, the uniform
discretization technique is used. This setting allows us to develop
a minimization of the (discrete) master constraint of the system
using a variational method. In order to accomplish this we need to
construct the Fock vacuum of the matter sector. We get this vacuum
solving the Hamilton equations for an effective Hamiltonian, which
turn out to be a Sturm-Liouville differential equation. It has to be
pointed out that we did not manage to find an exact solution for
this equation in the case we are interested in. Nevertheless, a
suitable approximate solution is provided. Once the trial state is
established, it is possible to perform the variational method in
order to minimize the master constraint. Finally, we get an
expression for the principal order of the expectation value of the
master constraint with respect to the lattice spacing $\epsilon$ and
the Planck length $\ell_p$. We found that, at least at the main
order, the expectation value of the master constraint does not
depend on the classical Schwarzschild radius.

At this point it is worth comparing our results with the ones of
Gambini, Pullin and Rastgoo. First, we have, at the leading order in
the expectation value of the master constraint the same situation as
in the cited paper, where the problem was worked out around a flat
Minkowskian spacetime. As in that case, we also have a diverging
master constraint in the continuum limit, but the theory gives a
good approximation to general relativity for small values of the
lattice separation. The reason for that, clearly explained in
\cite{Gambini09}, is due to the gauge fixing procedure employed to
get rid of the diffeormorphism constraint at the classical level.

This work can be extended in several ways. For example, performing a
polymeric quantization for the coupled system, or trying to work out
this computation avoiding the gauge fixing of the diffeomorphism
constraint. Nevertheless, these options would clearly rise the
technical complexity of the problem. Anyway, in our opinion, this is
a promising research line. A deeper understanding of this setting
will provide us with a suitable vacuum state in this context opening
the possibility of performing a detailed study of the Hawking
radiation within the LQG framework. We left this problem for future
investigation.


\section*{Acknowledgements}

The contents of this work are part of the Master Thesis of Eckhard
Strobel. The authors want to thank Alexander Stottmeister for
several enlightening discussions. We also thank Rodolfo Gambini, 
Jorge Pullin and Thomas Thiemann for the careful reading of the manuscript and their
comments to improve it.

This work was in part supported by the Spanish MICINN research
grants FIS2008-01980 and FIS2009-11893. IG is supported by the
Department of Education of the Basque Government under the
``Formaci\'{o}n de Investigadores'' program.


\appendix

\section{Computation of expectation values for basic operators}

In this appendix we give the expressions for several basic
expectation values that are used along the paper.

First, we compute the expectation value of the operator
$\hat{E}^{\varphi}(i)$:
\[
\langle \Psi_{\vec{\sigma}}|\hat{E}^{\varphi}(i)|\Psi_{\vec{\sigma}}
\rangle=\int_{-\infty}^{\infty}{d\vec{\mu}\,\ell_p^2\mu_i  \prod_{j}
\sqrt{\frac{2}{\pi \sigma(j)}} \exp\left(-\frac{2}{\sigma(j)}
\left(\mu_j-\frac{r_1(j)\epsilon}{\ell_p^2}\right)^2\right)}
=\epsilon r_1(i)\,,
\]
where $|\Psi_{\vec{\sigma}}\rangle$ is the gravitational part of the
trial state given by the equation (\ref{eq:trial}).

\bigskip

For the expectation value of trigonometric functions we proceed in
the same way, and find
\begin{eqnarray}
\langle \Psi_{\vec{\sigma}}|\cos(2\rho\hat{K}_{\varphi}(i)
)|\Psi_{\vec{\sigma}}\rangle&=&\exp\left(-\frac{2\rho^2}{\sigma(i)}
\right)\,,
\label{eq:exp11}\\
\langle \Psi_{\vec{\sigma}}|\cos(4\rho\hat{K}_{\varphi}(i)
)|\Psi_{\vec{\sigma}}\rangle&=&\exp\left(-\frac{8\rho^2}{\sigma(i)}
\right)\,,\\
\label{eq:Sin/E}
\langle \Psi_{\vec{\sigma}}|\frac{1}{\sqrt{\hat{E}^\varphi(i)}}
\sin(\rho\hat{K}_{\varphi}(i) )\frac{1}{\sqrt{\hat{E}^\varphi(i)}}
|\Psi_{\vec{\sigma}}\rangle&=&0\,,\\
\langle \Psi_{\vec{\sigma}}|\frac{1}{(\hat{E}^\varphi(i))^{3/2}}
\sin(\rho\hat{K}_{\varphi}(i) )\frac{1}{(\hat{E}^\varphi(i))^{3/2}}
|\Psi_{\vec{\sigma}}\rangle&=&0\,,\\
\langle
\Psi_{\vec{\sigma}}|\frac{1}{\sqrt{\hat{E}^\varphi(i)}}
\sin(3\rho\hat{K}_{\varphi}(i)
)\frac{1}{\sqrt{\hat{E}^\varphi(i)}}|\Psi_{\vec{\sigma}}\rangle&=&0\,.
\label{eq:exp12}
\end{eqnarray}

\bigskip

Now we use the inverse \((\hat{E}^\varphi)^{-3/2}\) operator,
following the prescription given for LQC in \cite{Bojowald:2001vw,
Ashtekar06}. We obtain for the eigenvalue of $|\vec{\mu}\ra$:
\begin{align}
(\hat{E}^\varphi(i))^{-3/2}| \vec{\mu} \rangle=
\ell_p^{-2}\left(\frac{2}{3\rho}\right)^6\left(
(\mu_i+\rho)^{3/4}-(\mu_i-\rho)^{3/4}\right)^6|
\vec{\mu} \rangle\,.
\end{align}
This can be used to calculate expectation values of operators
involving the inverse of \(\hat{E}^\varphi\):
\begin{align}
\langle \Psi_{\vec{\sigma}}|\frac{1}{(|\hat{E}^\varphi(i)|)^k}
|\Psi_{\vec{\sigma}}\rangle=\langle
\Psi_{\vec{\sigma}}|\left(|\hat{E}^\varphi(i)|\right)^{m-k}
\left(\frac{1}{(|\hat{E}^\varphi(i)|)^{3/2}}\right)^{2m/3}
|\Psi_{\vec{\sigma}}\rangle,
\end{align}
where \(m\ge k>0\) can be chosen arbitrarily, depending on the
prescription taken for applying the Thiemann's trick. In fact, there
is the same kind of ambiguity in LQC for the expression of inverse
volume operators.

We now concentrate on lattice spacings\, \(r\epsilon  \gg
\ell_p^2\). In this regime we obtain the following approximation
\begin{align}
\langle\Psi_{\vec{\sigma}}|\frac{1}
{|\hat{E}^\varphi(i)|^k}|\Psi_{\vec{\sigma}}\rangle&\approx
\frac{1}{\epsilon^k r_1(i)^k}+\frac{1}{\epsilon^{k+2}
r_1(i)^{k+2}}\left(\frac{k(k+1)}{8} \sigma(i)+\frac{5m}{24}\rho^2
\right)\, \ell_p^4\,. \label{eq:1/Ek}
\end{align}
As commented before, the dependence of (\ref{eq:1/Ek}) on \(m\)
shows an ambiguity which occurs due to the use of the inverse
operators. More specifically, for the inverse operators needed in
our case we obtain:
\begin{align}
\langle\Psi_{\vec{\sigma}}|\frac{1}
{(\hat{E}^\varphi(i))^2}|\Psi_{\vec{\sigma}}\rangle
&=\frac{1}{\epsilon^2r_1(i)^2}+\frac{1}{\epsilon^4 r_1(i)^4}
\alpha(i)\, \ell_p^4+\mathcal{O}\left(\frac{\ell_p^8}
{\epsilon^6}\right)\,, \label{eq:1/E2}
\\
\langle\Psi_{\vec{\sigma}}|\frac{1}
{(\hat{E}^\varphi(i))^4}|\Psi_{\vec{\sigma}}
\rangle&=\frac{1}{\epsilon^4r_1(i)^4}+\frac{1}{\epsilon^6
r_1(i)^6}\beta(i) \,\ell_p^4
+\mathcal{O}\left(\frac{\ell_p^8}{\epsilon^8}\right)\,,
\end{align}
with
$$
\alpha(i)=\left(\frac{5 m_\alpha}{24 }\rho^2
+\frac{3}{4}\sigma(i)\right)\,,\qquad \beta(i) =\left(\frac{5
m_\beta}{24 }\rho^2 +\frac{5}{2}\sigma(i)\right),
$$
where \(m_\alpha \geq 2\) and \(m_\beta \geq 4\). Analogously one
gets
\begin{align}
\label{eq:exp22} \langle
\Psi_{\vec{\sigma}}|\frac{1}{{\hat{E}^\varphi}(i)}
\cos(2\rho\hat{K}_{\varphi}(i))\frac{1}{{\hat{E}^\varphi}(i)}
|\Psi_{\vec{\sigma}}\rangle&=\exp\left(-\frac{2\rho^2}{\sigma(i)}
\right)\left(\frac{1}{\epsilon^2r_1(i)^2}+ \frac{1}{\epsilon^4
r_1(i)^4}\gamma(i) \,\ell_p^4\right)
+\mathcal{O}\left(\frac{\ell_p^8}{\epsilon^6}\right)\,,
\end{align}
with
$$
\gamma(i)=\left(\left(1+\frac{5
m_\gamma}{12}\right)\rho^2+\frac{3}{4}\sigma(i)\right)\,, \qquad
m_\gamma \geq 1.
$$

\section{Solution to the Sturm-Liouville problem}
\label{app:SL}

\begin{figure}[htbp]
   \centering
    \includegraphics[width=0.5\textwidth]{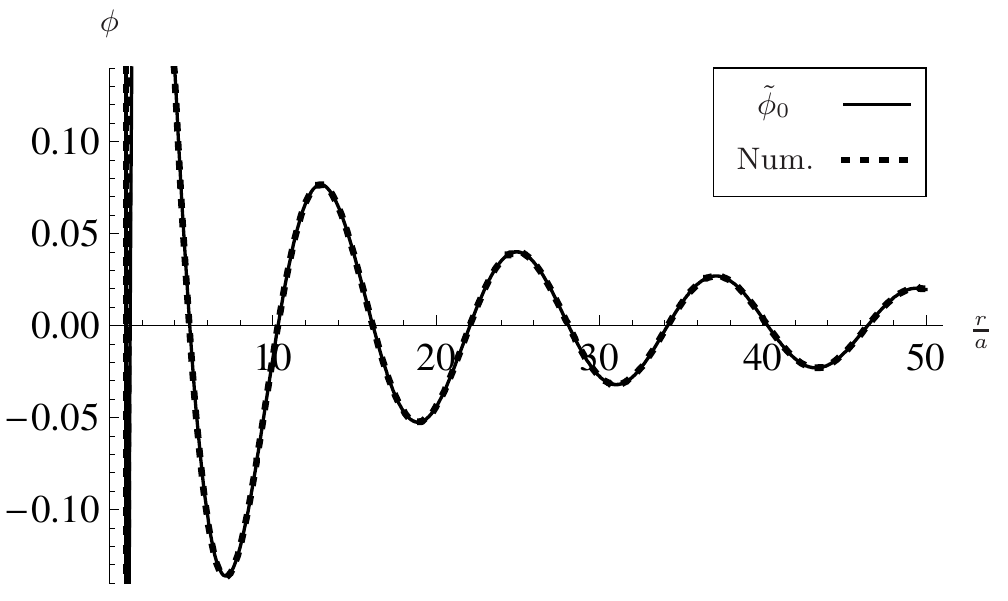}
    \caption{Comparison of the approximate solution
    \(\tilde{\phi}_0(r,\omega)\) (solid line) given by equation
    (\ref{eq:asol1}) with the numerical one (dashed line) for $\omega=L/a$.
    Notice that they overlap completely, showing the quality of
    the approximation.}
       \label{fig:SL1}
\end{figure}

\begin{figure}[htbp]
    \centering
        \includegraphics[width=0.5\textwidth]{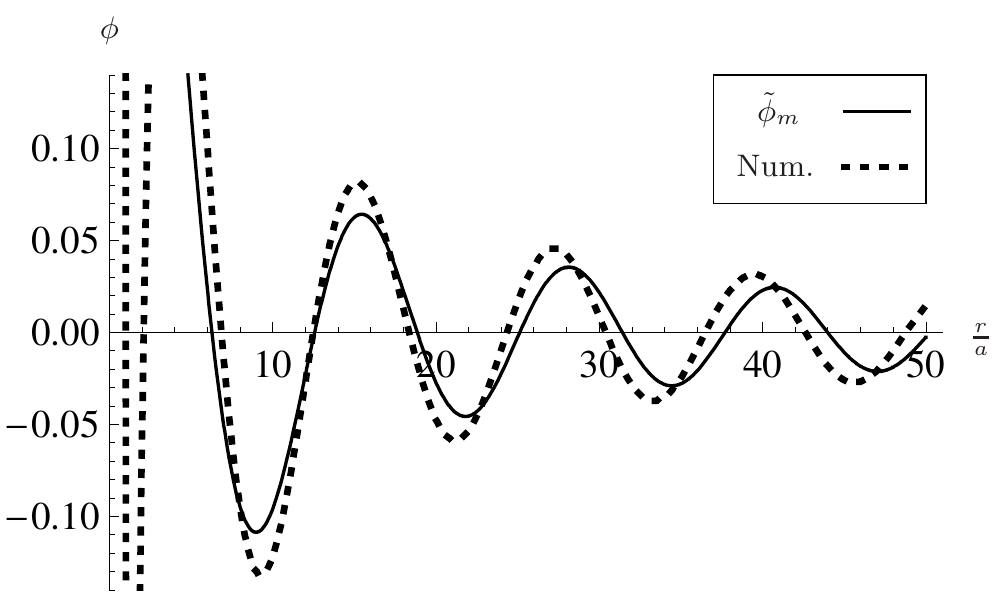}
        \caption{Comparison of the Minkowskian solution
    \(\tilde{\phi}_m(r,\omega)\) (solid line) given by the equation
    (\ref{eq:minksol}) with the numerical one (dashed line) for $\omega=L/a$.
    Notice that \(\tilde{\phi}_m(r,\omega)\) is not a
    suitable approximation.}
       \label{fig:SL2}
\end{figure}

We consider the Sturm-Liouville problem (\ref{eq:SL}) for the
leading order $\phi_0(r,\omega)$ in the expansion of the solution
in powers of \(\ell_p^4/\epsilon^2\).
\begin{equation}
\left(2 L \,\left(1-\frac{a}{r}\right)  r^2
\phi_0'(r,\omega)\right)'+\frac{\omega^2}{2L}
\frac{r^2}{1-\frac{a}{r}} \phi_0(r,\omega)=0\,, \label{eq:SL0}
\end{equation}
where $L=(1-2\Lambda)$. We were able to find the function
\begin{equation}
\tilde{\phi}_0(r,\omega)=\frac{1}{r}\sin\left(\frac{\omega}{2L}
\,r^\ast \right)\,, \label{eq:asol1}
\end{equation}
where
$$
r^\ast=r+a\log\left(\frac{r}{a}-1\right)
$$
is the usual tortoise coordinate of the external region of the
Schwarzschild metric. The approximate solution (\ref{eq:asol1})
fulfills
\begin{equation}
\left(2L  \,\left(1-\frac{a}{r}\right)  r^2
\tilde{\phi}_0'(r,\omega)\right)'+\frac{\omega^2}{2L}
\frac{r^2}{1-\frac{a}{r}} \tilde{\phi}_0(r,\omega)=-2\frac{a}{r}L
\tilde{\phi}_0(r,\omega)\,,
\end{equation}
which for \(r\gg a\) properly approximates the Sturm-Liouville
equation (\ref{eq:SL0}). We can observe in figure \ref{fig:SL1} that
indeed the approximation is very accurate (it overlaps completely
the exact numerical solution). Another possible option could be
motivated by the fact that in the regime $r\gg a$ the equation
(\ref{eq:SL0}) becomes the Sturm-Liouville problem of the Minkowski
case, studied in \cite{Gambini09}, with exact solution
\be
\tilde{\phi}_m(r,\omega)=\frac{1}{r}\sin\left(\frac{\omega}{2L}
r\right)\,.
\label{eq:minksol}
\ee
However this Minkowskian solution does not succeed in approximating
the exact numerical solution, as it is illustrated by figure
\ref{fig:SL2}. We will therefore work with
\(\tilde{\phi}_0(r,\omega)\).
As mentioned before solutions of (\ref{eq:SL0}) should fulfill the
orthogonality relation (\ref{eq:ortho}). We can use this fact to
normalize our solution, which is then given by
\begin{equation}
\tilde{\phi}_0(r,\omega)=\sqrt{\frac{2}{\pi}}\frac{1}{r}\sin
\left(\frac{\omega}{2L}\,r^\ast\right)\,. \label{eq:asol}
\end{equation}

\section{Coefficient operators and expectation values}
\label{app:coeff}

\subsection{Coefficient operators}

The operator form of the coefficients of the discrete master
constraint (\ref{eq:disHam}) is:
{\allowdisplaybreaks
\begin{align}
\hat{c}_{11}&=\frac{1}{\hat{E}^\varphi(i)^4}\,,
\\[10pt]
\hat{c}_{12}&=\frac{1}{(\hat{E}^\varphi(i))^{3/2}}
\frac{\sin(\rho\hat{K}_{\varphi}(i) )}{\rho}
\frac{1}{(\hat{E}^\varphi(i))^{3/2}}\,,
\\[10pt]
\hat{c}_{13}&=-\frac{1}{\hat{E}^\varphi(i)^2}\,,
\\[10pt]
\hat{c}_{22}&=\frac{1}{2\rho^2}\left(\frac{1}{\hat{E}^\varphi(i)^2}
-\frac{1}{\hat{E}^\varphi(i)}\cos(2\rho\hat{K}_{\varphi}(i) )
\frac{1}{\hat{E}^\varphi(i)}\right)\,,
\\[10pt]
\hat{c}_{23}&=-\frac{1}{(\hat{E}^\varphi(i))^{1/2}}\frac{\sin(\rho\hat{K}_{\varphi}(i) )}{\rho}\frac{1}{(\hat{E}^\varphi(i))^{1/2}}\,,
\\[10pt]
\hat{c}_{33}&=1\,,
\\[10pt]
\begin{split}
\hat{c}_{14}&=\left(r(i)\frac{1}{2\rho^2}-r(i+1)
\frac{1-\cos(2\rho\hat{K}_{\varphi}(i+1))}{2\rho^2}
-(1-2\Lambda)\epsilon\right)\frac{1}{(\hat{E}^\varphi(i))^2}
\\
&-r(i)\frac{1}{\hat{E}^\varphi(i)}
\frac{\cos(2\rho\hat{K}_{\varphi}(i))}{2\rho^2}
\frac{1}{\hat{E}^\varphi(i)}+\frac{r(i+1)^3\epsilon^2}
{\hat{E}^\varphi(i+1)^2\hat{E}^\varphi(i)^2}
-\frac{r(i)^3\epsilon^2}{\hat{E}^\varphi(i)^4}\,,
\end{split}
\\[10pt]
\begin{split}
\hat{c}_{24}&=\left(\frac{r(i+1)^3\epsilon^2}{\hat{E}^\varphi(i+1)^2}
-r(i+1)\frac{1-\cos(2\rho\hat{K}_{\varphi}(i+1))}{2\rho^2}
-(1-2\Lambda)\epsilon\right)\frac{1}{(\hat{E}^\varphi(i))^{1/2}}
\frac{\sin(\rho\hat{K}_{\varphi}(i) )}{\rho}
\frac{1}{(\hat{E}^\varphi(i))^{1/2}}
\\
&+r(i)\frac{1}{\hat{E}^\varphi(i)^{1/2}}\frac{3\sin(\rho
\hat{K}_{\varphi}(i))-\sin(3\rho\hat{K}_{\varphi}(i))}{4\rho^3}
\frac{1}{\hat{E}^\varphi(i)^{1/2}}-r(i)^3\epsilon^2\frac{1}
{(\hat{E}^\varphi(i))^{3/2}}\frac{\sin(\rho\hat{K}_{\varphi}(i)
)}{\rho}\frac{1}{(\hat{E}^\varphi(i))^{3/2}}\,,
\end{split}
\\[10pt]
\begin{split}
\hat{c}_{34}&=r(i+1)\frac{1-\cos(2\rho\hat{K}_{\varphi}(i+1))}
{2\rho^2}-r(i)\frac{1-\cos(2\rho\hat{K}_{\varphi}(i))}{2\rho^2}
-\frac{r(i+1)^3\epsilon^2}{\hat{E}^\varphi(i+1)^2}+\frac{r(i)^3\epsilon^2}{\hat{E}^\varphi(i)^2}+(1-2\Lambda)\epsilon\,,
\end{split}
\\[10pt]
\begin{split}
\hat{c}_{44}&=(1-2\Lambda)^2\epsilon^2+2(1-2\Lambda)\epsilon\left(r(i+1)\frac{1-\cos(2\rho
\hat{K}_{\varphi}(i+1))}{2\rho^2}-r(i)\frac{1-\cos(2\rho
\hat{K}_{\varphi}(i))}{2\rho^2}\right)
\\
&+r(i+1)^2\frac{3-4\cos(2\rho\hat{K}_{\varphi}(i+1))+\cos(4\rho
\hat{K}_{\varphi}(i+1))}{8\rho^4}-2r(i)r(i+1)\frac{1-\cos(2\rho
\hat{K}_{\varphi}(i))}{2\rho^2}\frac{1-\cos(2\rho
\hat{K}_{\varphi}(i+1))}{2\rho^2}\\
&+r(i)^2\frac{3-4\cos(2\rho\hat{K}_{\varphi}(i))+\cos(4\rho
\hat{K}_{\varphi}(i))}{8\rho^4}+2\left(r(i)\frac{1-\cos(2\rho
\hat{K}_{\varphi}(i))}{2\rho^2}-\frac{r(i+1)}{2\rho^2}-(1-2\Lambda)\epsilon\right)
\frac{r(i+1)^3\epsilon^2}{\hat{E}^\varphi(i+1)^2}
\\
&+2\left(r(i+1)\frac{1-\cos(2\rho\hat{K}_{\varphi}(i+1))}
{2\rho^2}-\frac{r(i)}{2\rho^2}+(1-2\Lambda)\epsilon\right)\frac{r(i)^3\epsilon^2}
{\hat{E}^\varphi(i)^2}
+2r(i)^4\epsilon^2\frac{1}{\hat{E}^\varphi(i)}\frac{\cos(2\rho
\hat{K}_{\varphi}(i))}{2\rho^2}\frac{1}{\hat{E}^\varphi(i)}
\\
&+2r(i+1)^4\epsilon^2\frac{1}{\hat{E}^\varphi(i+1)}
\frac{\cos(2\rho\hat{K}_{\varphi}(i+1))}{2\rho^2}
\frac{1}{\hat{E}^\varphi(i+1)}+\frac{r(i+1)^6\epsilon^4}{\hat{E}^\varphi(i+1)^4}
-2\frac{r(i)^3\epsilon^2}{\hat{E}^\varphi(i)^2}
\frac{r(i+1)^3\epsilon^2}{\hat{E}^\varphi(i+1)^2}
+\frac{r(i)^6\epsilon^4}{\hat{E}^\varphi(i)^4}\,.
\end{split}
\end{align}
}
Using (\ref{eq:exp11})-(\ref{eq:exp12}) and
(\ref{eq:1/E2})-(\ref{eq:exp22}) one can now calculate the
expectation values of the coefficients $\langle \hat{c}_{ij}
\rangle=\langle\Psi_{\vec{\sigma}}^{\text{trial}}|\hat{c}_{ij}
|\Psi_{\vec{\sigma}}^{\text{trial}}\rangle
=\langle\Psi_{\vec{\sigma}}|\hat{c}_{ij}|\Psi_{\vec{\sigma}}\rangle$.
We remark that
\begin{equation}
\langle \hat{c}_{12} \rangle=\langle \hat{c}_{23} \rangle=\langle \hat{c}_{24} \rangle=0\,.
\label{eq:van}
\end{equation}

\subsection{Expectation values of the ``matter Hamiltonians''}

Let us focus now on the expectation values of the expressions of the
discrete master constraint which contain ``matter Hamiltonians''.
Since the coefficients (\ref{eq:van}) vanish, we only need to
compute $\langle (\hat{H}^{(1)}_{\text{matt}}(i))^2\rangle$,
$\langle (\hat{H}^{(2)}_{\text{matt}}(i))^2\rangle$ and  $\langle
\hat{H}^{(1)}_{\text{matt}}(i)\rangle$. The continuum version of
(\ref{eq:H1}) can be written as
\begin{equation}
\hat{H}^{(1)}_{\text{matt}}=\epsilon^2 \ell_p^2\left(
\hat{P}^\phi(r,t)^2+ r^4\hat{\phi}'(r,t)^2\right)\,.
\end{equation}
Using \(\langle \hat{H}^{(j)}_{\text{matt}}(i)\rangle=\langle
0|\hat{H}^{(j)}_{\text{matt}}(i)|0\rangle\),\,\, \(\langle
0|\hat{C}(\omega)\hat{C}^\dagger(\omega')|0\rangle=\delta(\omega-\omega')\)
and that all the other combinations of the \(\hat{C}\)-operators
vanish, one gets:
\begin{align}
\langle\hat{H}^{(1)}_{\text{matt}}\rangle&=\frac{\epsilon^2
\ell_p^2}{2}\left(A(r)^2  \int_0^\infty d\omega  \, \omega
(\tilde{\phi}_0(r,\omega))^2+ r^4 \int_0^\infty d\omega \frac{1}{
\omega } (\tilde{\phi}'_0(r,\omega))^2\right)\,.
\end{align}
For $(\hat{H}^{(1)}_{\text{matt}})^2$\,, with the given factor
ordering we obtain
\begin{equation}
(\hat{H}^{(1)}_{\text{matt}})^2=\epsilon^4 \ell_p^4\left(
\hat{P}^\phi(r,t)^4+2 r^4\hat{P}^\phi(r,t)\hat{\phi}'(r,t)^2
\hat{P}^\phi(r,t)+r^8\hat{\phi}'(r,t)^4\right)\,. \label{eq:H12}
\end{equation}
The expectation value of (\ref{eq:H12}) becomes
\begin{align}
\langle(\hat{H}^{(1)}_{\text{matt}})^2\rangle=\frac{\epsilon^4\ell_p^4}
{4}\left(3A(r)^4I_1^2+2A(r)^2r^4\left(2I_3+I_1 I_2\right) +3 r^8
I_2^2\right)\,,
\end{align}
where
\begin{align}
\label{eq:I1} I_1=\int_0^\infty d\omega\,
\omega(\tilde{\phi}_0(r,\omega))^2\,,\qquad I_2=\int_0^\infty
d\omega \frac{1}{ \omega } (\tilde{\phi}'_0(r,\omega))^2\,\qquad
I_3=\int_0^\infty d\omega
\,\tilde{\phi}_0(r,\omega)\tilde{\phi}'_0(r,\omega).
\end{align}
Now, for \((\hat{H}^{(2)}_{\text{matt}})^2\), which following from
(\ref{eq:H2}) has the continuum limit
\begin{equation}
(\hat{H}^{(2)}_{\text{matt}})^2=r^2 \ell_p^4\epsilon^2
\hat{P}^\phi(r,\omega) \hat{\phi}'(r,\omega)^2
\hat{P}^\phi(r,\omega)\,,
\end{equation}
we obtain
\begin{equation}
\langle(\hat{H}^{(2)}_{\text{matt}})^2\rangle=r^2 A(r)^2
\epsilon^2\ell_p^4\left(2I_3+I_1 I_2\right)\,.
\end{equation}
Summarizing and going back to the discrete theory we get the
following expectation values for the ``matter Hamiltonians'':
\begin{align}
\langle\hat{H}^{(1)}_{\text{matt}}(i)\rangle&=\frac{\ell_p^2}{2}
\epsilon^3\left(A(r(i))^2 I_1(i)+ r(i)^4 I_2(i)\right)\,,
\\
\langle(\hat{H}^{(1)}_{\text{matt}}(i))^2\rangle&=
\frac{\ell_p^4}{4}\epsilon^6\left(3A(r(i))^4I_1(i)^2+3 r(i)^8
I_2(i)^2+2A(r(i))r(i)^4\left(2I_3(i)^2+I_1(i)
I_2(i)\right)\right)\,,
\\
\langle(\hat{H}^{(2)}_{\text{matt}}(i))^2\rangle&= r(i)^2 A(r(i))^2
\epsilon^4\ell_p^4\left(2I_3(i)^2+I_1(i) I_2(i)\right)\,,
\\
\langle\hat{H}^{(3)}_{\text{matt}}(i)\rangle&=2 \Lambda \epsilon\,,
\\
\langle(\hat{H}^{(3)}_{\text{matt}}(i))^2\rangle&=4 \Lambda^2 \epsilon^2\,,
 \\
\langle\hat{H}^{(1)}_{\text{matt}}\hat{H}^{(3)}_{\text{matt}}(i)\rangle&=
\Lambda l_p^2\epsilon^4\left(A(r(i))^2 I_1(i)+r(i)^4 I_2(i)\right)\,,
\end{align}
where we used \(\rho_\text{vac}=2\Lambda/\ell_p^2\).

Now all we need to calculate to get all the components of the master
constraint (\ref{eq:dismas}) is the three integrals (\ref{eq:I1}).
Using the approximate solution (\ref{eq:asol}) and taking into
account the discretization of the integral one gets:
\begin{align}
I_1&=\frac{2}{\pi}\frac{1}{r^2}\left(
\frac{\pi^2}{\epsilon^2}+\frac{L^2}{2(r^\ast)^2}
-\frac{\cos(\frac{2\pi}{\epsilon L}r^\ast)}{2(r^\ast)^2}L^2
-\frac{\sin(\frac{2\pi}{\epsilon L}r^\ast)}{\epsilon\,r^\ast}\pi
L\right) \,, \label{eq:I11}
\\
\begin{split}
I_2&=\frac{1}{\pi}\frac{1}{r^4}\left(\gamma-\text{Ci}\left(\frac{2\pi}{\epsilon
L}r^\ast\right) +\log\left(\frac{2\pi}{\epsilon
L}r^\ast\right)\right)
-\frac{2}{\pi}\frac{1}{r^3}\frac{1}{1-\frac{a}{r}}
\frac{\sin^2\left(\frac{\pi}{\epsilon L}r^\ast\right)}{r^\ast}\\
&+\frac{1}{2\pi L^2} \frac{1}{r^2\left(1-\frac{a}{r}\right)^2}
\left(\frac{\pi^2}{\epsilon^2}-\frac{L^2}{2(r^\ast)^2}
+\frac{\cos(\frac{2\pi}{\epsilon L}r^\ast)}{2(r^\ast)^2}L^2
+\frac{\sin(\frac{2\pi}{\epsilon L}r^\ast)}{\epsilon\,r^\ast}\pi
L\right)\,, \label{eq:I21}
\end{split}
\\
I_3&=\frac{1}{\pi L}\frac{1}{r^2}\frac{1}{1-\frac{a}{r}}
\left(\frac{\sin(\frac{2\pi}{\epsilon L}r^\ast)}{2(r^\ast)^2}L^2
-\frac{\cos(\frac{2\pi}{\epsilon L}r^\ast)}{\epsilon\,r^\ast}\pi
L\right)
-\frac{2}{\pi}\frac{1}{r^3}\left(\frac{\pi}{\epsilon}-\frac{\sin(\frac{2\pi}{\epsilon
L}r^\ast)}{2r^\ast}L\right) \,,\label{eq:I31}
\end{align}
where $\text{Ci}(x)$ is the cosine integral
function\footnote{$\text{Ci}(x)\equiv\gamma+\log x+\int_0^xdt(\cos
t-1)\,.$}\, and $\gamma$ is the Euler's constant.





\end{document}